\documentclass[preprint,prd,showpacs,showkeys]{revtex4}
\usepackage{graphicx}
\usepackage{color}
\usepackage[fleqn]{amsmath}
\usepackage{amssymb, graphics}
\usepackage{multirow}
\usepackage{latexsym}
\usepackage{bm}
\usepackage{wrapfig}
\usepackage{fancybox}
\def\be{\begin{equation}}
\def\ee{\end{equation}}
\def\bea{\begin{eqnarray}}
\def\eea{\end{eqnarray}}
\begin{document}

\title{Thermodynamic Geometry Of Charged Rotating BTZ Black Holes}

\author{M. Akbar$^1$, H. Quevedo$^{2,3}$, K. Saifullah$^4$, A. S\'anchez$^5$,
and Safia Taj$^{1,3}$} \affiliation{$^1$Center for Advanced
Mathematics and Physics, National University of Sciences and
Technology, H-12, Islamabad, Pakistan\\
$^2$Instituto de Ciencias
Nucleares, Universidad Nacional Aut\'onoma de M\'exico,
 AP 70543, M\'exico, DF 04510, Mexico\\
 $^3$ICRANet, Dipartimento di Fisica, Universit\`a di Roma "La
Sapienza",  I-00185 Roma, Italy\\$^4$Department of Mathematics,
Quaid-i-Azam University, Islamabad, Pakistan \\$^5$Departamento de
Posgrado, CIIDET, AP 752, Quer\'etaro, QRO 76000, Mexico}

\begin{abstract}
We study the thermodynamics and the thermodynamic geometries of charged
rotating BTZ (CR-BTZ) black holes in (2+1)-gravity. We investigate the
thermodynamics of these systems within the context of the Weinhold and Ruppeiner
thermodynamic geometries and the recently developed
formalism of geometrothermodynamics (GTD).
Considering the behavior of the heat capacity and the Hawking temperature,
we show that Weinhold and Ruppeiner geometries cannot describe completely
the thermodynamics of these black holes and of their limiting case of vanishing
electric charge. In contrast, the Legendre invariance imposed on the metric
in GTD allows one to describe the CR-BTZ black holes and their limiting cases
in a consistent and invariant manner.


\end{abstract}

\maketitle

\section{Introduction}
\label{intro}

An important characteristic of a black hole consists of its
thermodynamic features: a Hawking temperature proportional to its
surface gravity on the horizon and an entropy proportional to its
horizon area \cite{haw,bek}, satisfying the first law of black hole
thermodynamics \cite{bch}. However, it is still a challenging
problem to find the statistical origin of black hole thermodynamics.
As one knows, the divergence of the heat capacity is an indication
of a second-order phase transition in ordinary thermodynamic
systems. Using this fact, Davies \cite{dav} argued that phase
transitions appear in black hole thermodynamics and the phase
transition point is the one where the heat capacity diverges
\cite{hut}.

On the other hand, several authors have proposed
to include geometric concepts in order to investigate
the properties of the equilibrium space of thermodynamic systems.
For instance,
Weinhold \cite{wnd} introduced in the equilibrium space a Riemannian
metric defined in terms of the second derivatives of the internal energy $U$ with
respect to entropy and other extensive variables of a thermodynamic
system. Weinhold metric is given by
\begin{equation}
g^{W}_{ij}=\partial_{i}\partial_{j}U(S,N^{r}).
\label{wei}
\end{equation}
However, the geometry based on this metric seems to be
meaningless in the context of pure equilibrium thermodynamics. In
1979, Ruppeiner \cite{rup} introduced a Riemannian metric structure
in thermodynamic fluctuation theory, and related it to the
second derivatives of the entropy. This geometric structure was used
to find the significance of the distance between equilibrium states
and to study the thermodynamics of equilibrium systems.
It was observed by Ruppeiner \cite{rupp,rupnr} that in thermodynamic
fluctuation theory the Riemannian curvature of the Ruppeiner metric
measures the complexity of the underlying statistical mechanical
model. Ruppeiner metric is defined as
\begin{equation}
g^{R}_{ij}=\partial_{i}\partial_{j}S(M,N^{r}),
\label{rup}
\end{equation}
where $S $ is the entropy, $U $ denotes the energy and $N^{r} $ are
other extensive variables of the system. The Ruppeiner geometry is
conformally related to the Weinhold geometry \cite{mru,sal} as
\begin{equation}
ds^{2}_{R}=\frac{1}{T}ds^{2}_{W},
\label{rwr}
\end{equation}
where $T$ is the temperature of the system under consideration.
Eq.(3) often provides a more convenient way to
compute the Ruppeiner metric.

One of the aims of the application of geometry in thermodynamics is
to describe phase transitions in terms of curvature singularities and
to interpret curvature  as a measure of thermodynamic
interaction. Since the proposal of Weinhold, many investigations
have been carried out to understand the thermodynamic geometry
of various thermodynamic systems. The Weinhold and Ruppeiner
geometries have been analyzed in a number of black hole families
to study phase space, critical behavior, and stability properties
\cite{ferr,jjkb,jjka,am,ama,aman,scws,cc,sst,med,mz}. In some
particular cases, it was found that Weinhold and  Ruppeiner geometries carry
information about the phase transitions structure. In fact, this is true in the
case of the ideal gas, whose curvature vanishes, and the van der
Waals gas for which the thermodynamic curvature
becomes singular at those points where phase transitions
occur. Unfortunately, the obtained results are contradictory in the case
of black holes. For instance, for the Kerr black hole Weinhold
metric predicts no phase transitions at all \cite{am} whereas
Ruppeiner metric, with a very specific thermodynamic potential,
predicts phase transitions which are compatible with the results of
standard black hole thermodynamics \cite{scws}. Nevertheless, a
change of the thermodynamic potential affects the Ruppeiner geometry in such a
way that the resulting curvature singularity does not correspond to a
phase transition.
Another example is provided by the Ba\~nados-Teitelboim-Zanelli (BTZ) black hole
thermodynamics for which  the curvature of the
equilibrium space turns out to be flat \cite{cc,sst,med}. This
flatness is usually interpreted as a consequence of the lack of
thermodynamic interaction. However, if one applies an invariant
approach the resulting manifold is curved \cite{qs}.

Recently, the formalism of geometrothermodynamics (GTD) was
developed in order to unify in a consistent manner the geometric
properties of the phase space and the space of equilibrium states
\cite{qjmp,qgrg,qv}. Legendre invariance plays an important role in
this formalism. It has been shown that there exist thermodynamic
metrics that correctly describe the thermodynamic behavior of the
ideal and the van der Waals gas. In fact, for the ideal gas the
curvature vanishes whereas for the van der Waals gas the curvature
is non-zero and diverges only at those points where phase
transitions take place. Moreover, there exists a thermodynamic
metric with non-vanishing curvature which correctly describes the
thermodynamic properties of  black holes \cite{all}.

In fact, the problem of using Weinhold or Ruppeiner metrics in
equilibrium space is that the results can depend on the choice of
thermodynamic potential, i. e., the results are not invariant with
respect to Legendre transformations \cite{sib,mnss}. These results
indicate that, in the case of black holes, geometry and
thermodynamics are compatible only for a specific thermodynamic
potential. However, it is well known that ordinary thermodynamics
does not depend on the thermodynamic potential, i.e., it is
invariant with respect to Legendre transformations. GTD incorporates
Legendre invariance into the geometric structures of the phase space
and equilibrium space so that the results do not depend on the
choice of thermodynamic potential. The phase transition structure
contained in the heat capacity of black holes \cite{davi} becomes
completely integrated in the scalar curvature of the Legendre
invariant metric so that a curvature singularity corresponds to a
phase transition.

In the present work we investigate the Weinhold and Ruppeiner
geometries of charged rotating BTZ (CR-BTZ) black holes. We show
that both geometries are curved, indicating the presence
of thermodynamic interaction. However, in the limiting case of
vanishing electric charge certain inconsistencies appear.
We also use
GTD to derive a Legendre invariant metric for the CR-BTZ black holes. Our
main result is that GTD correctly describes the thermodynamics
of the CR-BTZ black holes and that no inconsistencies appear in the
limiting cases of vanishing
electric charge.

This paper is organized as follows. In Section \ref{sec:crbtz}, we
review some known facts about CR-BTZ black holes and present the
main features of their thermodynamics. In Section \ref{sec:wr}, we
examine Weinhold and Ruppeiner geometries of the CR-BTZ black hole.
We review and apply 3-dimensional GTD in Section \ref{sec:gtd}.
Finally, the last Section contains the conclusions. Throughout this
paper we use the units in which $c=\hbar=8G=1$.


\section {The charged rotating BTZ black hole}
\label{sec:crbtz}

The charged rotating BTZ  black hole solutions \cite{bhtz,mtz} in $(2
+ 1)$ spacetime dimensions are particular solutions to the field equations derived
from the action \cite{mtz,ach}
\begin{equation}
I =\frac{1}{2\pi } \int dx^{3}\sqrt{-g} \left(R +
2\Lambda-\frac{\pi}{2}F_{\mu \nu}F^{\mu \nu}\right)\ .
\end{equation}%
The Einstein field equations are given by
\begin{equation}
G_{\mu \nu}-\Lambda g_{\mu \nu}=\pi T_{\mu \nu} ,
\end{equation}%
where $T_{\mu \nu}$ is the energy-momentum tensor of the
electromagnetic field:
\begin{equation}
T_{\mu \nu} = F_{\mu \rho}F_{\nu \sigma}g^{\rho \sigma}-\frac{1}{4}
g_{\mu \nu}F^{2}.
\end{equation}%
The corresponding line element for the CR-BTZ solution is
\begin{equation}
ds^{2}=-f(r)dt^{2}+\frac{dr^{2}}{f(r)}+r^{2}\left(d\phi-\frac{J}{2r^{2}}
dt\right)^{2}\ ,
\end{equation}%
with lapse function:
\begin{equation}
f(r)=-M+\frac{r^{2}}{l^{2}}+\frac{J^{2}}{4r^{2}}-\frac{\pi }{2}Q^{2}\ln r \ .
\end{equation}
Here $M$ and $J$ are the mass and angular momentum respectively,
and $Q$ is the charge carried by the black hole.  The horizons of the
CR-BTZ metric correspond to the roots of the lapse function $f(r)$. Depending on
these roots there are three cases for the CR-BTZ configuration
\cite{aa,cms}: Two distinct roots $r_\pm$ determine the standard
CR-BTZ black holes; two repeated real roots correspond to a
single horizon that determines an extreme black hole; the absence of real roots
implies that no horizon exists and the configuration corresponds to that of
a naked singularity. We shall investigate the first case in this work.

In terms of the exterior
horizon radius $r_+$, the black hole
mass and the angular momentum are given respectively by
\begin{equation}
M=\frac{r_{+}^{2}}{l^{2}}+\frac{J^{2}}{4 r_{+}^{2}}-\frac{\pi}{2}Q^2
\ln(r_{+}),
\label{feqm}
\end{equation}%
and
\begin{equation}
J=2r_{+} \sqrt{M-\frac{r_{+}^{2}}{l^{2}}-\frac{\pi}{2}Q \ln(r_{+})},
\label{10}
\end{equation}%
with the corresponding angular velocity on the horizon
\begin{equation}
\Omega=2r_{+}^{2}\frac{\partial M}{\partial J}%
\bigg|_{r=r_{+}}=\frac{J}{2r_{+}^{2}}=\frac{1}{r_{+}}
\sqrt{M-\frac{r_{+}^{2}}{l^{2}}-\frac{\pi}{2}Q \ln(r_{+})}.
\label{11}
\end{equation}%
The Hawking temperature $T$ at the black hole horizon is
\begin{equation}
T=\frac{1}{4\pi}\frac{df}{dr}=\frac{1}{4 \pi}\left(
\frac{2r_{+}}{l^{2}}-\frac{J^{2}}{2
r_{+}^{3}}-\frac{\pi}{2}\frac{Q^{2}} {r_{+}}\right),
\label{temp}
\end{equation}%
and the electric potential is given by
\begin{equation}
\phi=\frac{\partial M}{\partial Q}\bigg|_{r=r_{+}}=-\pi Q\ln r_{+}.
\label{13}
\end{equation}%
Furthermore, using the fundamental postulate of black hole thermodynamics,
the entropy of the CR-BTZ black hole is defined as
\begin{equation}
S =4\pi r_{+}.
\label{entropy}
\end{equation}%
In terms of this entropy, the corresponding thermodynamic
fundamental equation and the temperature for the CR-BTZ black hole
are given respectively by
\begin{equation}
M=\left(\frac{S}{4\pi l}\right)^2+\left(\frac{2\pi
J}{S}\right)^2-\frac{\pi Q^{2}}{2}\ln \frac {S}{4\pi},
\label{feq}
\end{equation}%
and
\begin{equation}
T=\left(\frac{\partial M}{\partial S}\right)_{J,Q} =\frac{S}{8\pi^{2}l^{2}}-%
\frac{8\pi^{2}J^{2}}{S^{3}}-\frac{\pi Q^{2}}{2S}.
\label{temp1}
\end{equation}
The thermodynamic quantities $T,S,J,Q$ and $M$ obey the first law of
thermodynamics
\begin{equation}
dM =T dS +\Omega dJ +\phi dQ.
\label{flaw0}
\end{equation}

An important quantity for the analysis of the thermodynamic properties is the
heat capacity of the CR-BTZ black hole \cite{aa},
$C_{J,Q}=\left(
\partial M/\partial T\right) _{J,Q},$ which is given by
\begin{equation}
C_{J,Q}={S}\frac{S^{4}-4\pi ^{3}l^{2}Q^{2}S^{2}-64\pi
^{4}l^{2}J^{2}}{S^{4}+4\pi ^{3}l^{2}Q^{2}S^{2}+192\pi
^{4}l^{2}J^{2}}, \label{heatc0}
\end{equation}%
or using the horizon radius $r_{+}$ as coordinate, by
\begin{equation}
C_{J,Q}=4\pi r_{+} \frac{4r_{+}^{4}-\pi l^{2}Q^{2}r_{+}^{2}-l^{2}J^{2}%
}{4r_{+}^{4}+\pi l^{2}Q^{2}r_{+}^{2}+3l^{2}J^{2}} . \label{heatc}
\end{equation}%

\begin{figure}
\includegraphics[width=5cm]{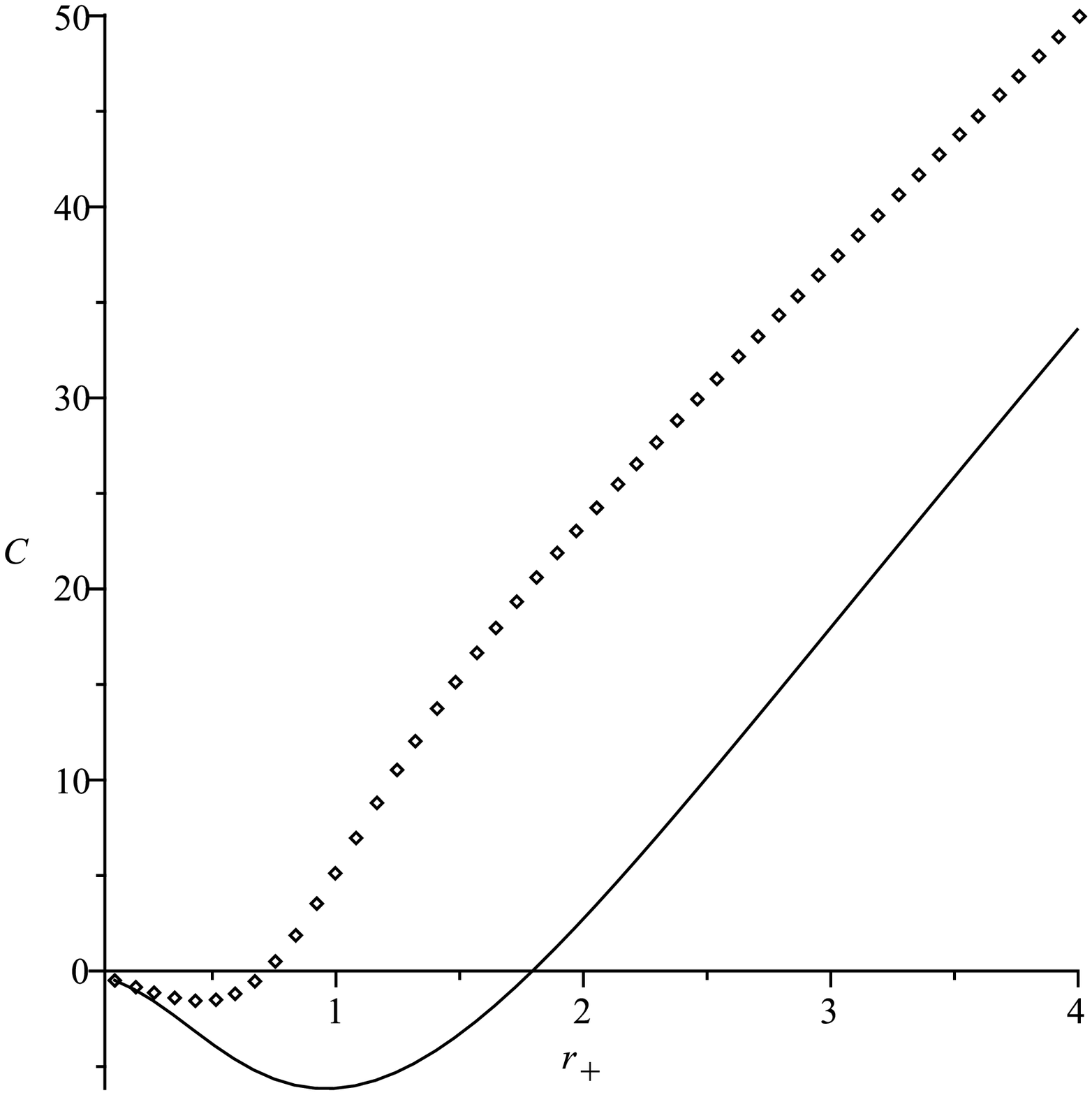}
\includegraphics[width=5cm]{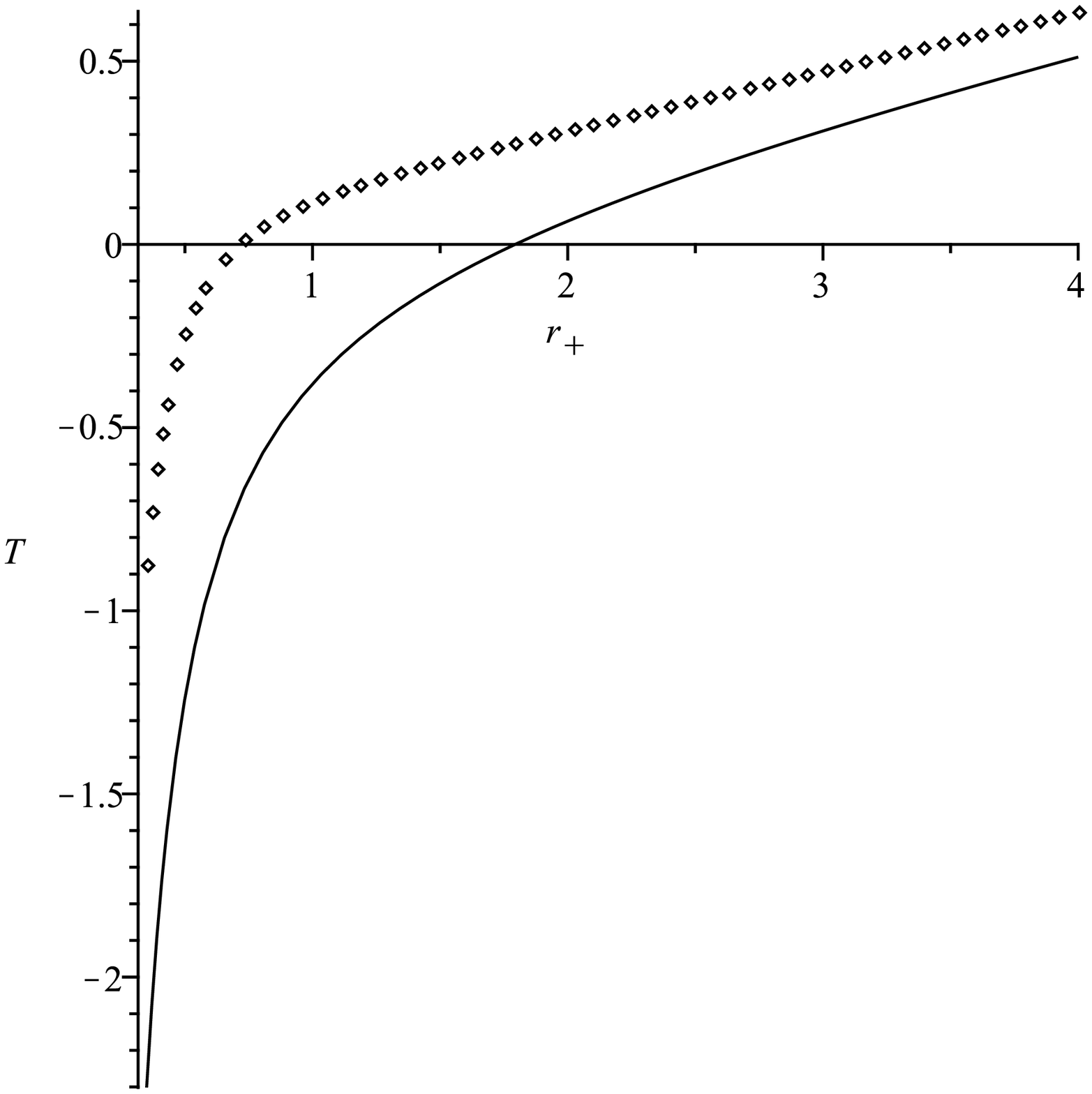}
\caption{Behavior of the heat capacity and temperature as  functions
of the event horizon radius $r_{+}$ of a CR-BTZ black hole with
$Q=2$, $J=1$ and $l=1$. Temperature and heat capacity vanish at $r_+\approx 1.79$.
The unstable region $(C_{J,Q}<0)$
corresponds to an unphysical region with negative temperature. To
illustrate the contribution of the charge we include the plots
(dotted curves) for the case of the non-static BTZ black hole with
$Q=0$ \cite{qs,cai}.}
\label{figone}
\end{figure}

From expressions (\ref{temp}) and (\ref{entropy})  we conclude that the condition
\begin{equation}
S^{4}-4\pi ^{3}l^{2}Q^{2}S^{2}-64\pi ^{4}l^{2}J^{2} > 0 \label{cond}
\end{equation}
must be satisfied in order for the temperature to be positive
definite, a requirement which follows from the standard laws of
black hole thermodynamics. From the above condition and the
expression (\ref{heatc0}), it follows that the heat capacity is
always positive definite. This is an important observation which
implies that a CR-BTZ black hole with a positive definite
temperature must be a thermodynamically stable configuration. In
fact, a change of sign of the heat capacity  is usually associated
with a drastic change of the stability properties of a thermodynamic
system; a negative heat capacity represents a region of instability
whereas the stable domain is characterized by a positive heat
capacity.

It is worth mentioning that the heat capacity is a regular function
for all real positive values of the exterior horizon radius. In
fact, the denominator of the expression (\ref{heatc}) is always
positive and, consequently, $C_{J,Q}$ is a regular function, except
in the pathological case where $S=J=Q=0$. On the other hand, in
black hole thermodynamics, divergences of the heat capacity are
associated with second-order phase transitions. This implies that a
CR-BTZ black hole cannot undertake a phase transition associated
with a divergence of the heat capacity. The above observations
demonstrate that the CR-BTZ black hole is a completely stable
thermodynamic system with no phase transition structure. In this
work, we will use this fact in order to test different geometric
descriptions of the thermodynamics of the CR-BTZ black hole. Figure
\ref{figone} shows for selected values of $J$, $Q$ and $l$ the
behavior of heat capacity and temperature for a charged black hole
(solid line) and for a neutral black hole (dotted line). The
comparison of both curves shows that the charge essentially
increases the value of the horizon radius at which the heat capacity
and the temperature vanish. As the value of the horizon radius
increases the contribution of the charge decreases. Finally, for
very large values of the horizon radius the heat capacities and the
temperatures coincide, indicating that the contribution of the
charge is negligible.

\begin{figure}
\includegraphics[width=8cm]{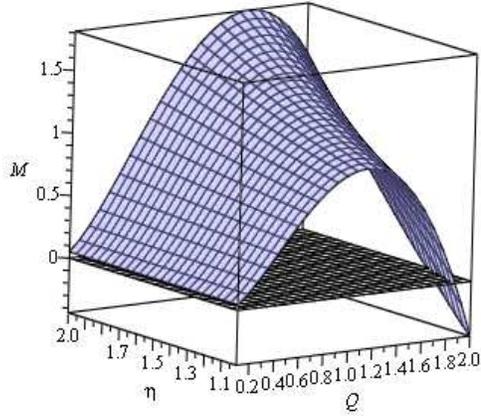}
\caption{The mass $M$ of an extreme CR-BTZ black hole with horizon
radius $r_*=\pi l^2 Q^2 (1+\eta)/8$ as a function of the charge $Q$
and the angular parameter $\eta>1$.   Here, we chose $l=1$ as a
representative value. The plane $M=0$ is plotted to visualize the
region where $M>0$. }
\label{figextr}
\end{figure}

One would expect that the limiting case $T\rightarrow 0$ or,
equivalently, $C\rightarrow 0$ corresponds to an extreme black hole
with only one horizon of radius, say, $r_*$. To analyze this
question it is necessary to find the domain of parameters for which
the equation $f(r)=0$ allows only one positive real root, and to
calculate the value of $T$ for this domain.  However, the equation
$f(r)=0$ cannot be solved analytically because of the presence of
the logarithmic term $\ln r$. An alternative procedure consists in
solving the equation $T=0$ for $r^2$ to obtain \be r^2:= r_*^2 =
\frac{\pi l^2Q^2}{8} (1+\eta)\ , \quad \eta =
\sqrt{1+\frac{16J^2}{\pi^2 l^2 Q^4}} \ , \ee and introducing this
solution into the equation $f(r)=0$ to obtain the value of the mass
at this radius, i. e., \be M= \frac{\pi Q^2}{4} \left[\eta -
\ln(1+\eta) -\ln\frac{\pi l^2Q^2}{8}\right]\ , \ee where we replaced
$J^2$ by using the definition of the auxiliary parameter $\eta$. Now
the question is whether the last expression represents a physical
mass, i.e. whether it is positive. A numerical analysis shows that
for any value $\eta>1$, a condition that follows from the definition
of $\eta$, there always exists a combination of values for $Q$ and
$l$ such that $M$ is positive. Figure \ref{figextr} shows an example
of the behavior of the mass for a fixed value of the parameter $l$.
We conclude that the limit $T\rightarrow 0$ indeed corresponds to an extreme
black hole.


\section {Weinhold and Ruppeiner geometries}
\label{sec:wr}

\begin{figure}
\includegraphics[width=5cm]{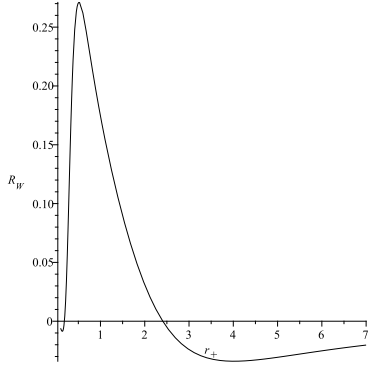}
\caption{Thermodynamic curvature for the Weinhold ($R_{W}$) geometry
as a function of the event horizon radius, $r_{+}$, of the CR-BTZ
black hole. Here, the free parameters are chosen as  $Q=2$, $J=1$
and $l=1$. The curvature is completely regular in the entire domain
of $r_{+}$. }
\label{figtwo}
\end{figure}

Now we construct the thermodynamic geometry of the CR-BTZ black hole
by using the Weinhold metric (\ref{wei}). In this case the extensive
variables are $N^r=\{J,Q\}$ so that the general Weinhold metric
becomes
\begin{equation}
\begin{split}
ds^{2}_{W}=& \left( \frac{\partial ^{2}M}{\partial S^{2}}\right)dS^{2}
+\left( \frac{\partial ^{2}M}{\partial J^{2}}\right)dJ^{2}
+\left( \frac{\partial ^{2}M}{\partial Q^{2}}\right)dQ^{2} \\&
+2\left(\frac{\partial^{2}M}{\partial S\partial J}\right)dSdJ+ 2\left(\frac{%
\partial^{2}M}{\partial J\partial Q}\right)dJdQ
+2\left(\frac{\partial^{2}M}{\partial Q\partial S}\right)dQdS,
\end{split}
\end{equation}
and in the special case of the CR-BTZ black hole we have
\begin{equation}
\begin{split}
ds^{2}_{W}= &
\left(\frac{1}{8\pi^{2}l^{2}}+\frac{24\pi^{2}J^{2}}{S^{4}}+
\frac{\pi Q^{2}}{2S^{2}}\right)dS^{2} +\frac{8\pi^{2}}{S^{2}}dJ^{2}
-\pi\ln\left(\frac{S}{4\pi}\right)dQ^{2} \\&
-\frac{32\pi^{2}J}{S^{3}}dSdJ-2\frac{\pi Q}{S}dSdQ\ .
\end{split}
\end{equation}
The corresponding scalar curvature is given by
\begin{equation}
R_W=\frac{l^2 r_+^2\Big[-4r_+^4(1+2\ln{r_+}+4\ln{r_+}^2)+\pi l^2
Q^2(9+2\ln{r_+}){{ r_+}}^{2} +J^2l^2(1+2\ln{r_+})¨\Big]}{\Big[
-4\,{{ r_+}}^{4}\ln{ r_+}  -\pi \,{Q}^{2}{l}^{2}( \ln{ r_+} +2){{
r_+}}^{2} +{J}^{2}{l}^ {2}\ln{ r_+}\Big] ^{2}} .
\end{equation}
The general behavior of the scalar curvature for the Weinhold geometry is
illustrated in Figure \ref{figtwo}.
We see that the thermodynamic curvature is regular for all positive
values of the horizon radius. At the value $r_+\approx 1.79$ (with
$l=1,\, J=1,\, Q=2)$, at which the temperature vanishes, the scalar
curvature is $R_W \approx 0.0527$. Moreover, it is positive and
regular in the interval $0.185 <r_+<1.79$, a region where the temperature
is negative, and in the interval $1.79 \leq r_+ < 2.4$, a region where the
temperature is positive.
This means that the Weinhold thermodynamic curvature
cannot differentiate between a CR-BTZ black hole with positive
temperature and a similar configuration with negative temperature.

We now investigate the limiting case of a vanishing charge. The
additional extensive variable in this case is $N^r=\{J\}$ so that
the Weinhold metric reduces to
\begin{equation}
ds^{2}_{W}=\left(\frac{1}{8\pi^{2}l^{2}}+\frac{24\pi^{2}J^{2}}{S^{4}} \right)dS^{2}
 -\frac{32\pi^{2}J}{S^{3}}dSdJ
+\frac{8\pi^{2}}{S^{2}}dJ^{2} \ ,
\end{equation}
and the corresponding scalar curvature becomes
\begin{equation}
R_W = 16\,{\frac {{\pi}^{2}{l}^{2}{S}^{6}}{ \left( {S}^{4}-64\,{\pi}^{4}{J
}^{2}{l}^{2} \right) ^{2}}}\ .
\end{equation}
We see that there exists a true curvature singularity at the value
$S^4=64\,{\pi}^{4} {J}^{2}{l}^{2}$  that, according to
Eq.(\ref{cond}) with $Q=0$, corresponds to the the limit of
vanishing temperature or, equivalently, to the extreme black hole
limit. This result shows that the Weinhold thermodynamic curvature
in this case correctly describes the transition from a region with
positive and well-defined temperature to a region with an unphysical
negative temperature. This is in contrast to what we obtained in the
case of a charged black hole  in which the Weinhold thermodynamic
curvature is not able to recognize the transition to an extreme
black hole with zero temperature.

\begin{figure}
\includegraphics[width=5cm]{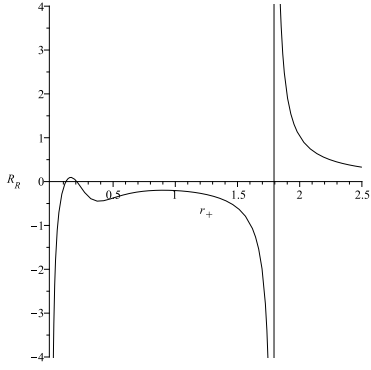}
\caption{Thermodynamic curvature, $R_{R}$, of the Ruppeiner geometry
as a function of the event horizon, $r_{+}$, for a CR-BTZ black hole
with $J=l=1$, $Q=2$. The only singularity is located at $r_{+}\approx 1.79$.}
\label{figthree}
\end{figure}

Let us now consider the Ruppeiner geometry. A direct computation of
the Ruppeiner metric (\ref{rup}) cannot be performed because it is
not possible to rewrite explicitly the fundamental equation
(\ref{feq}) in the the form $S=S(M,J,Q)$. Nevertheless, if we assume
the invariance of the line element under a change of thermodynamic
potential, the relationship (\ref{rwr}) can be used to derive the
Ruppeiner metric from the Weinhold metric. Then, we obtain
\begin{equation}
ds_R^2 =  \frac{dS^2}{S} +\frac{\pi}{T}\left[ \frac{8\pi}{S^2}\left(\frac{2J}{S}dS
-dJ\right)^2 +\left(\frac{Q}{S}dS-dQ\right)^2
-\left(1+\ln\frac{S}{4\pi}\right) dQ^2\right] \ .
\end{equation}
The corresponding thermodynamic curvature scalar $R_{R}$ turns out
to be nonzero, i.e., the space of its thermodynamic equilibrium
states is non-flat. The explicit form of $R_{R}$ cannot be written
in a compact form. Therefore, we perform a numerical analysis of its
behavior and the result is illustrated  in Figure \ref{figthree}.

The singularity located at $r_+\approx 1.79$ represents the limit
for which the heat capacity vanishes and the temperature becomes
negative. This shows that the Ruppeiner thermodynamic curvature
describes correctly the behavior of the CR-BTZ black hole.

In the limiting case of a vanishing charge, it is possible to
rewrite the fundamental equation (\ref{feqm}) as $S=S(M,J)$ in the
following manner
\begin{equation}
S=4\pi r_+ = \pi
\sqrt {8{M{l}^{2}} \left( 1+\sqrt {1-{\frac {{J}^{2}}{{M}^{
2}{l}^{2}}}} \right) }
\end{equation}
so that the Ruppeiner metric can be computed by using the definition (\ref{rup}).
Then
\begin{equation}
ds^2_R = -\frac{\pi l^2}{(r_+^2-r_-^2)^3} \left[ r_+(r_+^2 +
3r_-^2)(l^2 dM^2 + dJ^2) - 2lr_- (3r_+^2 + r_-^2) dM dJ\right]\ ,
\end{equation}
where
\begin{equation}
r_\pm^2  =\frac{l^2 M}{2} \left( 1 \pm \sqrt{ 1- \frac{J^2}{l^2 M^2} }\right)\ .
\end{equation}
A straightforward calculation shows that the curvature of this
metric vanishes identically, indicating the absence of thermodynamic
interaction, i.e., the thermodynamic variables $M$ and $J$ do not
generate thermodynamic interaction. This is a  peculiar result
because, as we have seen above,  the Ruppeiner geometry correctly
describes the thermodynamic behavior of the  CR-BTZ black hole. This
implies that only the charge $Q$ acts as a source of thermodynamic
interaction in the Ruppeiner geometry.  It seems that there is no
specific reason for the existence of this difference between
thermodynamic variables of this particular black hole configuration.


\section{Geometrothermodynamics of the CR-BTZ black hole}
\label{sec:gtd}

In order to describe a thermodynamic system with $n$ degrees of
freedom, we consider in GTD \textit{the thermodynamic phase space}
which is defined mathematically as a Riemannian contact manifold
$({\cal T} ,\Theta,G)$, where ${\cal T}$ is a $(2n+1)-$dimensional
manifold, $\Theta$ defines a contact structure on ${\cal T}$ and $G$
is a Legendre invariant metric on ${\cal T}$. The \textit{space of
equilibrium states} is an $n$-dimensional Riemannian manifold
$({\cal E}, g)$, where ${\cal E}\subset {\cal T}$ is defined by a
smooth mapping $\varphi:{\cal E}\rightarrow{\cal  T}$ such that the
pullback $ \varphi^*(\Theta)=0 $ and a Riemannian structure $g$ is
induced naturally in ${\cal E}$  by means of $g=\varphi^*(G)$.
It is
then expected in GTD that the physical properties of a thermodynamic
system in a state of equilibrium can be described in terms of the
geometric properties of the corresponding space of equilibrium
states ${\cal E}$ \cite{qjmp}.

To be more specific we introduce in the phase space ${\cal T}$ the
set of independent coordinates $Z^{A}=(\Phi,E^{a}, I^{a})$ with
$A=0, ...,2n$ and $a = 1, ..., n$, where $\Phi$ represents the
thermodynamic potential, and $E^{a}$ and $I^{a}$ are the extensive
and intensive thermodynamic variables, respectively. Consider the
Legendre-invariant Gibbs 1-form
\begin{equation}
\Theta_{G} = d\Phi - \delta_{ab}I^{a}dE^{b},    \ \  \delta_{ab}={\rm
diag}(1,...,1).
\end{equation}
The pair $({\cal T},\Theta)$ is called a contact manifold
\cite{herm} if ${\cal T}$ is differentiable and $\Theta$ satisfies
the condition $\Theta\wedge(d\Theta)^{n}\neq0$. Legendre invariance
\cite{arn} guarantees that the geometric properties of $G$ do not
depend on the thermodynamic potential used in its construction.

The smooth mapping $\varphi: {\cal E}
\rightarrow{\cal T}$ is given in terms of coordinates as $\varphi:
\{E^{a}\}\mapsto \{Z^{A}\} = \{\Phi(E^{a}),E^{a}, I^{a}(E^{a})\}$.
Consequently, the condition  $ \varphi^*(\Theta)=0 $ can be written
as the expressions
\begin{equation}
\frac{\partial \Phi}{\partial E^{a}}=\delta_{ab}I^{b},\ \  d\Phi
=\delta_{ab}I^{a}dE^{b}.
\end{equation}
which in ordinary thermodynamics correspond to the first law of
thermodynamics and the conditions for thermodynamic equilibrium,
respectively. In thermodynamics $\phi(E^{a})$ is known as the
fundamental equation from which all the equations of state of the
system can be derived \cite{call}. The second law of thermodynamics
is implemented in GTD as the convexity condition
$\partial ^{2}\Phi/\partial E^{a}\partial E^{b}\geq 0$. Furthermore,
the Euler and Gibbs-Duhen identities can be expressed as
$\Phi=\delta_{ab} I^a E^b$ and $\delta_{ab}E^a dI ^b =0$, respectively.

For the geometric description of the thermodynamics of the CR-BTZ
black hole in GTD, we first introduce the 7-dimensional phase space
${\cal T}$ with coordinates $M, S, J,Q, T,\Omega$ and $\phi$, a
contact 1-form
\begin{equation}
\Theta =dM-TdS-\Omega dJ-\phi dQ,
\end{equation}
which satisfies the condition $\Theta\wedge (d\Theta)^{3}\neq0$, and a
Legendre invariant metric
\begin{equation}
G=(dM-TdS-\Omega dJ-\phi dQ)^2+ TS (-dTdS+d\Omega
dJ+d\phi dQ).
\end{equation}
This particular metric is a special case of a metric used in \cite{qs}
to describe the region of positive temperature of the BTZ black hole.

Let ${\cal E}$ be a 3-dimensional subspace of ${\cal T}$ with coordinates
$E^{a} = ( S,Q, J)$, $a = 1, 2, 3,$ defined by means of a smooth
mapping $\varphi : {\cal E}\rightarrow {\cal T}$. The subspace
${\cal E}$ is called the space of equilibrium states if
$\varphi^{*}(\Theta)= 0$, where $\varphi^{*}$ is the pullback of
$\varphi$. Furthermore, a metric structure $g$ is naturally induced
on ${\cal E}$ by applying the pullback on the metric $G$
of ${\cal T}$ , i.e., $g=\varphi^{*}(G)$. It is clear that the condition
$\varphi^{*}(\Theta)=0$ leads immediately to the first law of
thermodynamics of black holes as given in Eq.(\ref{flaw0}). It also implies
the existence of the fundamental equation $M = M(S,Q, J)$ and the
conditions of thermodynamic equilibrium Eqs. (\ref{11})-(\ref{13}). Moreover, the
induced metric
\begin{equation}
g =\varphi^{*}(G) = S\frac{\partial M}{\partial S}
\left(-\frac{\partial^2 M} {\partial S^2}%
dS^{2}+\frac {\partial^2 M}{\partial J^2}dJ^2+\frac{\partial^2 M}{\partial Q^2}%
dQ^2\right)
\end{equation}
determines all the geometric properties of the equilibrium space
${\cal E}$. In the above expression we used the Euler identity to
simplify the form of the conformal factor. In order to obtain the
explicit form of the metric it is only necessary to specify the
thermodynamic potential $M$ as a function of $S$, $J$ and $Q$ as
given in Eq.(\ref{feq}). Another advantage of the use of GTD is that
it allows us to easily implement different thermodynamic
representations of the fundamental equation, given as $M = M(S,Q,
J),S = S(M,Q, J), Q = Q(S,M, J)$ or $J = J(S,M,Q)$ and redefine the
coordinates in ${\cal T}$ and the smooth mapping $\phi$ in such a
way that the condition $\varphi^{*}(\Theta)=0$ generates on ${\cal
E}$ the corresponding fundamental equation in the $S-$, $Q-$, or the
$J$-representation, respectively. The results obtained with
different representations of the same fundamental equation are
completely equivalent.

For the CR-BTZ black hole, using the
fundamental equation $M = M(S,J,Q)$  given in Eq.(\ref{feq}), the
thermodynamic metric can be written as
\begin{equation}
g ={\frac {{S}^{4}-64\,{\pi }^{4}{J}^{2}{l}^{2}-4\,{\pi }^{3}l^2{Q}^{2}{S}^{2}
}{8 {\pi }^{2}
{l}^{2}S^2}}
 \left[-\left(\frac{1}{8\pi^{2}l^{2}}+\frac{24\pi^{2}l^{2}}{S^4}+\frac{\pi Q^2%
}{2S^2}\right)dS^2+\frac{8\pi^{2}}{S^2}dJ^2-\pi\ln\frac{S}{4\pi}%
dQ^2\right].
\end{equation}
\begin{figure}
\includegraphics[width=5cm]{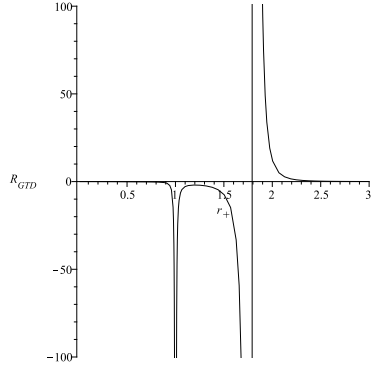} 
\caption{Thermodynamic curvature of the CR-BTZ black hole
($R_{GTD}$) as a function of event horizon $r_{+}$. A typical
behavior is depicted for the specific values $Q=2$, $J=l$ and
$l=1$.}
\label{figfour}
\end{figure}
The corresponding thermodynamic curvature turns out to be nonzero
and is given by
\begin{equation}
R_{GTD} = \frac{ 2l^4 r_+^4}{D_1 D_2}\left[ \frac{1}{\ln^2 r_+} +\frac{1}{D_1 D_2^2}
\left(\frac{N_1}{\ln r_+} +N_0  \right)\right]
\end{equation}
where
\be
D_1 = 4r_+^4 + \pi l^2Q^2 r_+^2 + 3 l^2 J^2\ , \quad D_2 = 4r_+^4 - l^2 J^2 - \pi
l^2Q^2 r_+^2 \nonumber\ ,
\ee
\be
\begin{split}
N_0 = & \ 4[ 6\,{J}^{6}{l}^{6}+23\,\pi \,{J}^{4}{l}^{6}{Q}^{2}{{ r_+^2}}+
 l^4 J^2\left( 15\,{\pi }^{2}{l}^{2}{Q}^{4}+8\,{J}^{2} \right)
{{ r_+^4}} + 4\pi l^4 Q^2 \left( 14\, \,{J}^{2}+{\pi }^{2}{Q
}^{4}{l}^{2} \right) {{ r_+^6}}  \\&  + 4 l^2 \left( {\pi }^{2}{Q}^{4}{l}^{2
}-40\,{J}^{2} \right) {{ r_+^8}}-16\,\pi \,{Q}^{2}{l}^{2}{
{ r_+^{10}}}+128\,{{ r_+^{12}}}] \nonumber\  ,
\end{split}
\ee
\be
\begin{split}
N_1 = & \ 2[ -15\,{J}^{6}{l}^{6}-6\,\pi \,{J}^{4}{l}^{6}{Q}^{2}{{ r_+^2}}
-3l^4 J^2 \left({\pi }^{2}{l}^{2}{Q}^{4}-20{J}^{2}
 \right) {{ r_+^4}}+128 \pi {J}^{2}{l}^{4}{Q}^{2}{{ r_+^6}} \\&
-4l^2 \left( 4 {J}^{2} - 5 {\pi }^{2}l^2{Q}^{4} \right) {{ r_+^8}}+96\,\pi
 \,{Q}^{2}{l}^{2}{{ r_+^{10}}}+64\,{{ r_+^{12}}} ]\nonumber\ .
\end{split}
\ee We see from the expression for the scalar curvature that the
curvature singularities can be situated at those values of the
parameters where $D_1=0$, $D_2=0$ or $\ln r_+ =0$. For real values
of the parameters the condition $D_1 = 4r_+^4 + \pi l^2Q^2 r_+^2 + 3
l^2 J^2=0$ cannot be satisfied. In fact, this term appears in the
denominator of the heat capacity (\ref{heatc}) and determines the
absence of phase transitions of the CR-BTZ black hole. The
singularities determined by the roots of the equation $D_2 = 4r_+^4
- l^2 J^2 - \pi l^2Q^2 r_+^2=0$ coincide with the points where $T=0$
or, equivalently, where the heat capacity (\ref{heatc}) vanishes.
This implies that the no physical region of negative temperatures is
isolated from the allowed region with positive temperatures by a
true curvature singularity. The third singularity located at $\ln
r_+=0$ can be interpreted as a critical point that is not determined
by the heat capacity (\ref{heatc}). In fact, at $r_+=1$ the second
derivative of the mass $\partial^2 M /\partial Q ^2 =0$, indicating
either the transition into a region of instability or a second order
phase transition. The singular behavior of the GTD scalar curvature
is illustrated in
 Figure \ref{figfour}.

\begin{figure}
\includegraphics[width=5cm]{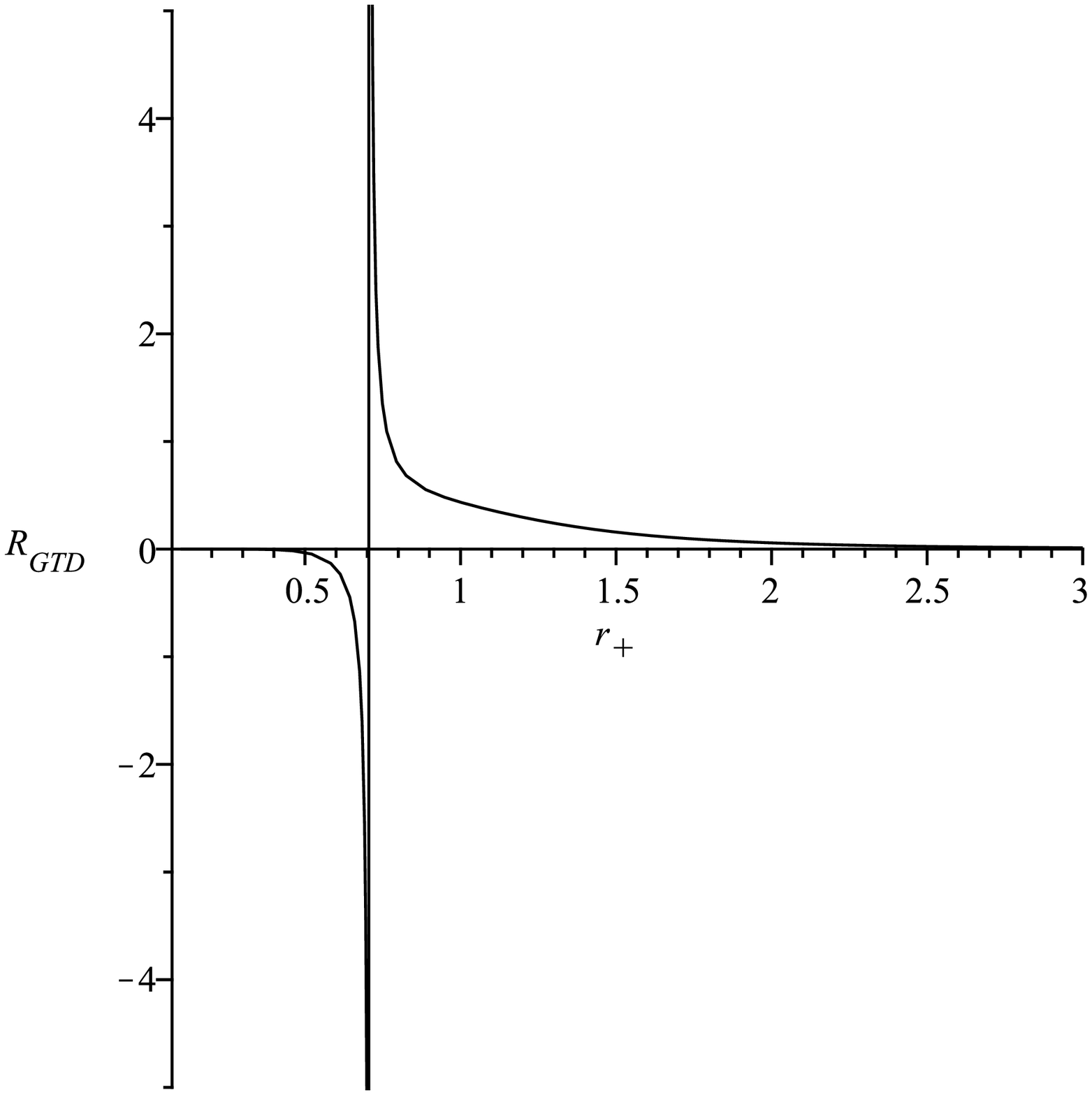}
\includegraphics[width=5cm]{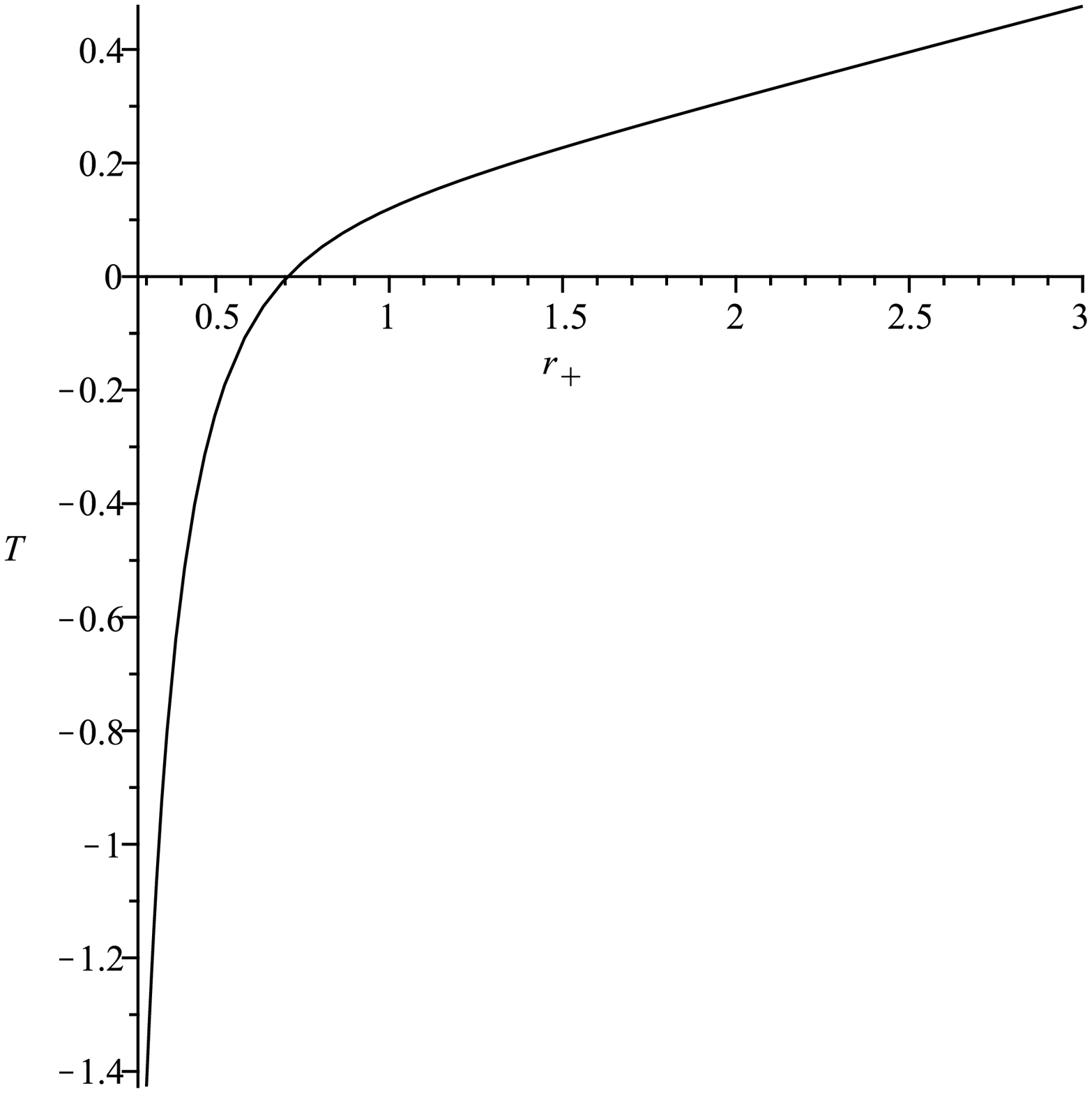}
\caption{Behavior of the scalar curvature in GTD and temperature as
functions of the event horizon radius $r_{+}=S/4\pi$ of a  neutral
rotating BTZ black hole with $J=1$ and $l=1$. The curvature
singularity coincides with the point of zero temperature.}
\label{figfive}
\end{figure}

Let us now consider the limiting case of vanishing charge. The
geometrothermodynamic metric reduces to \be g ={\frac
{{S}^{4}-64\,{\pi }^{4}{J}^{2}{l}^{2} }{8 {\pi }^{2} {l}^{2}S^2}}
\left[-\left(\frac{1}{8\pi^{2}l^{2}}+\frac{24\pi^{2}l^{2}}{S^4}
\right)dS^2+ \frac{8\pi^{2}}{S^2}dJ^2\right] , \ee and the
corresponding scalar curvature can be written as \be R_{GTD}= {\frac
{256 \, {l}^{4}{\pi }^{4}{S}^{8}}{ \left( {S}^{4}+192\,{\pi }^{4
}{J}^{2}{l}^{2} \right) ^{2} \left( {S}^{4}-64\,{\pi
}^{4}{J}^{2}{l}^ {2} \right) }} \ee The behavior of this scalar and
the temperature is depicted in Figure \ref{figfive}. It follows that
in general a curvature singularity appears when the condition
${S}^{4}-64\,{\pi }^{4}{J}^{2}{l}^{2} =0$ is satisfied which,
according to Eq.(\ref{temp1}) with $Q=0$, corresponds to a zero
temperature. We conclude that the invariant metric proposed in GTD
correctly describes the limiting case of a neutral BTZ black hole.

\section{Conclusions}

In this work, we analyzed the thermodynamics and the thermodynamic
geometry of the charged rotating Ba\~nados-Teitelboim-Zanelli
(CR-BTZ) black hole. By considering the behavior of the heat
capacity and the Hawking temperature, we found that this black hole
configuration is free of phase transitions and stable. In fact, the
instability region is characterized by a non-physical negative
temperature. Moreover, we performed a numerical analysis which shows
that in the limiting case of zero temperature the black hole becomes
extreme.

We analyzed the thermodynamic geometry based on the Weinhold metric and
found that the corresponding thermodynamic curvature is free of singularities
in the entire equilibrium manifold. This result is not in accordance with the
analysis of the behavior of the heat capacity and the Hawking temperature that
indicates the presence of an unphysical region with negative temperature
for $r_+\leq 1.79$ (the additional parameters are chosen as $l=1,\ J=1$ and $Q=2$).
We conclude that Weinhold geometry does not describe correctly
the thermodynamic geometry in this specific case. However, in the limiting case of
a vanishing electric charge there exists a true curvature singularity that is located
at the point where the temperature vanishes. It is not clear why the presence of an
electric charge cannot be handled correctly  in the context of the Weinhold
thermodynamic metric.

Although it is not possible to calculate explicitly the Ruppeiner
metric, it can be derived from the Weinhold metric by using a
conformal transformation with  the inverse of the temperature as the
conformal factor. A numerical analysis of the Ruppeiner
thermodynamic curvature shows that it is smooth and well-behaved in
the region $r_+>1.79$, with a true curvature singularity situated at
$r_+\approx 1.79$. This is exactly the value of the horizon radius
at which the Hawking temperature vanishes. We interpret this result
as indicating that the Ruppeiner geometry correctly describes the
thermodynamics of the CR-BTZ black hole.  However, in the limiting
case of vanishing electric charge the Ruppeiner metric turns out to
be flat. Since a vanishing thermodynamic curvature is usually
interpreted as indicating the absence of thermodynamic interaction,
it is not clear why Ruppeiner geometry correctly describes the
thermodynamics of the CR-BTZ black hole but fails in the limiting
neutral case.

Finally, we analyzed the properties of a Legendre invariant metric
proposed in the context of geometrothermodynamics (GTD). In this
case, the curvature can be calculated explicitly and it turns out
that it possesses a true curvature singularity at those points where
the Hawking temperature vanishes. In the entire region where the
CR-BTZ black hole corresponds to a stable thermodynamic system with
no phase transition structure, the thermodynamic curvature of GTD is
described by a smooth function of all the thermodynamic variables.
In the limiting case of vanishing electric charge, the metric
proposed in GTD is also able to correctly describe the thermodynamic
properties of the black hole configuration in the sense that it is
finite and smooth in the region where the black is stable, but
possesses a true curvature singularity at the point where the
temperature vanishes. Since the Weinhold and Ruppeiner metrics are
not invariant with respect to Legendre transformations, we conclude
that the Legendre invariance imposed in the context of GTD is an
important property to describe geometrically the thermodynamics of
black  holes without intrinsic contradictions.

\section*{Acknowledgments}

This work was partially supported by DGAPA-UNAM grant IN106110. Two
of us (HQ and ST) would like to thank ICRANet for support. Two of us
(KS and ST) gratefully acknowledge a research grant from the Higher
Education Commission of Pakistan under Project No: 20-638. We are
grateful to Prof. Asghar Qadir for interesting comments,
encouragement and support.



\begin{thebibliography}{99}

\bibitem{haw} S. W. Hawking, Commun. Math. Phys. \textbf{43}, 199 (1975).
\bibitem{bek} J. D. Bekenstein, Phys. Rev. D \textbf{7}, 2333 (1973).
\bibitem{bch} J. M. Bardeen, B. Carter, and S. W. Hawking, Commun. Math. Phys.
\textbf{31}, 161 (1973).
\bibitem{dav} P. C. W. Davies, Proc. Roy. Soc. Lond. A \textbf{353}, 499 (1977);
 P. C. W. Davies, Rep. Prog. Phys. 41, 1313(1977);
 P. C. W. Davies, Class. Quant. Grav. \textbf{6}, 1909 (1989).
\bibitem{hut} P. Hut, Mon. Not. R. Astron. Soc. \textbf{180}, 379 (1977).
\bibitem{wnd} F. Weinhold, J. Chem. Phys. \textbf{63}, 2479 (1975).
\bibitem{rup} G. Ruppeiner, Phys. Rev. A \textbf{20}, 1608 (1979).
\bibitem{rupp} G. Ruppeiner, Rev. Mod. Phys. \textbf{67}, 605 (1995).
\bibitem{rupnr} G. Ruppeiner, Rev. Mod. Phys. \textbf{68}, 313 (1996).
\bibitem{mru} R. Mrugala, Physica A \textbf{125}, 631 (1984).
\bibitem{sal} P. Salamon, J. D. Nulton, and E. Ihrig, J. Chem. Phys. \textbf{80},
436 (1984).
\bibitem{ferr} S. Ferrara, G. W. Gibbons, and R. Kallosh, Nucl. Phys. B \textbf{500},
75 (1997).
\bibitem{jjkb} D. A. Johnston, W. Janke, and R. Kenna, Acta Phys. Polon. B \textbf{34},
4923 (2003).
\bibitem{jjka} W. Janke, D. A. Johnston, and R. Kenna, Physica A \textbf{336}, 181
(2004).
\bibitem{am} J. E. \AA man, I. Bengtsson, and N. Pidokrajt, Gen. Rel. Grav.
\textbf{35,} 1733 (2003).
\bibitem{ama} J. E. \AA man and N. Pidokrajt, Phys. Rev. D \textbf{73}, 024017
(2006).
\bibitem{aman} J. E.  \AA man and N. Pidokrajt, Gen. Rel. Grav.\textbf{38},
1305 (2006).
\bibitem{scws} J. Shen, R. G. Cai, B. Wang, and R. K. Su, [gr-qc/0512035].
\bibitem{cc} R. G. Cai and J. H.Cho, Phys. Rev. D \textbf{60}, 067502 (1999).
\bibitem{sst} T. Sarkar, G. Sengupta, and B. N. Tiwari, J. High Energy Phys.
\textbf{0611} 015 (2006).
\bibitem{med} A. J. M. Medved, Mod. Phys. Lett. A \textbf{23}, 2149 (2008).
\bibitem{mz} B. Mirza and M. Zamaninasab, J. High Energy Phys. \textbf{06}, 059
(2007).
\bibitem{qs} H. Quevedo and A. S\'anchez, Phys. Rev. D \textbf{79}, 024012 (2009).
\bibitem{qjmp} H. Quevedo, J. Math. Phys. \textbf{48}, 013506 (2007).
\bibitem{qgrg} H. Quevedo, Gen. Rel. Grav. \textbf{40}, 971 (2008).
\bibitem{qv} H. Quevedo and A. V\'azquez, AIP Conf. Proc. \textbf{977}, 165 (2008);
arXiv:math-ph/0712.0868
\bibitem{all} J. L. \'Alvarez, H. Quevedo, and A. S\'anchez, Phys. Rev. D
\textbf{77}, 084004
(2008); H. Quevedo and A. S\'anchez, J. High Energy Phys.
\textbf{09}, 034 (2008); H. Quevedo and A. S\'anchez, Phys. Rev. D
\textbf{79}, 024012 (2009);  S. Taj and H. Quevedo,
Geometrothermodynamics of higher dimensional black holes in
Einstein-Gauss-Bonnet theory (2010, in preparation).
\bibitem{sib} P. Salamon, E. Ihrig, and R. S. Berry, J. Math. Phys. 24, 2515 (1983).
\bibitem{mnss} R. Mrugala, J. D. Nulton, J. C. Schon, and P. Salamon, Phys.
Rev. A \textbf{41}, 3156 (1990).
\bibitem{davi} P. C. W. Davies, Rep. Prog. Phys. \textbf{41}, 1313 (1978).
\bibitem{bhtz} M. Ba\~nados, M. Henneaux, C. Teitelboim, and J. Zanelli, Phys. Rev. D
\textbf{48} 1506 (1993).
\bibitem{mtz} C. Martinez, C. Teitelboim, and J. Zanelli, Phys. Rev. D \textbf{61}
104013 (2000).
\bibitem{ach} A. Achucarro and  M. E. Ortiz, Phys. Rev. D \textbf{48} 3600  (1993).
\bibitem{aa} M. Akbar and A. A. Siddiqui, Phys. Lett. B \textbf{656}, 217
(2007).
\bibitem{cms} M. Cadoni, M. Melis, and  M. R. Setare, Class. Quantum Grav. \textbf{25}
195022 (2008); M. Cadoni and C. Monni, Phys. Rev. D \textbf{80}
 024034 (2009); Y. S. Myung, Y. W. Kim, and  Y. J. Park,
arXiv:0903.2109v1 [hep-th].
\bibitem{cai} R. G. Cai and J. H. Cho, Phys. Rev. D \textbf{60} 067502 (1999).
\bibitem{herm} R. Hermann, (Marcel Dekker, New York, 1973); W. L. Burke, (Cambridge
University Press, Cambridge, UK, 1987).
\bibitem{arn} V. I. Arnold, Mathematical
methods of classical mechanics (Springer Verlag, New York, 1980).
\bibitem{call} B. Callen, Thermodynamics and an introduction to thermostatics (John
Wiley and Sons, Inc., New York, 1985).
\end{thebibliography}
\end{document}